\pdfoutput=1

\documentclass[11pt,twoside,a4paper,cmspaper,final,collab]{cms-tdr}
\def\svnVersion{302642}\def\cmsCernNo{2013-037}\def\cmsCernDate{\today}\def\cmsMessage{Published in Physics Letters B as \href{http://dx.doi.org/10.1016/j.physletb.2013.09.057}{\doi{10.1016/j.physletb.2013.09.057.}}}

\begin{document}\cmsNoteHeader{HIG-13-006}

\hyphenation{had-ron-i-za-tion}
\hyphenation{cal-or-i-me-ter}
\hyphenation{de-vices}

\RCS$Revision: 206511 $
\RCS$HeadURL: svn+ssh://svn.cern.ch/reps/tdr2/papers/HIG-13-006/trunk/HIG-13-006.tex $
\RCS$Id: HIG-13-006.tex 206511 2013-09-10 13:58:59Z alverson $
\newlength\cmsFigWidth
\ifthenelse{\boolean{cms@external}}{\setlength\cmsFigWidth{0.95\columnwidth}}{\setlength\cmsFigWidth{0.4\textwidth}}
\newlength\cmsFigWidthX
\ifthenelse{\boolean{cms@external}}{\setlength\cmsFigWidthX{0.95\columnwidth}}{\setlength\cmsFigWidthX{0.7\textwidth}}
\ifthenelse{\boolean{cms@external}}{\providecommand{\cmsLeft}{top}}{\providecommand{\cmsLeft}{left}}
\ifthenelse{\boolean{cms@external}}{\providecommand{\cmsRight}{bottom}}{\providecommand{\cmsRight}{right}}
\cmsNoteHeader{HIG-13-006} 
\title{Search for a Higgs boson decaying into  a \Z and a photon in \Pp\Pp\ collisions at $\sqrt{s}$ = 7 and 8\TeV}

\date{\today}

\abstract{
A search for a Higgs boson decaying into a \Z boson and a photon is described.
The analysis is performed using proton-proton collision datasets recorded by the CMS detector at the LHC.
Events were collected at  center-of-mass energies of 7\TeV and 8\TeV, corresponding to integrated luminosities of
5.0\fbinv and 19.6\fbinv, respectively.
The selected events are required to have
opposite-sign electron or muon pairs.
No excess above standard model predictions has been found  in the
120--160\GeV mass range
and the first limits on the Higgs boson production
cross section times the  $\PH\to\Z\gamma$  branching fraction at the LHC have been derived.
The observed limits are between about 4 and 25 times the
standard model cross section times the branching fraction.
The observed  and expected limits for $m_{\ell\ell\gamma}$ at  125\GeV are within one order of magnitude of
the standard model prediction.
Models predicting the Higgs boson production cross section times the $\PH\to\Z\gamma$ branching fraction
to be larger than one order of magnitude of the standard model prediction are excluded for most of the 125--157\GeV mass range.
}

\hypersetup{%
pdfauthor={CMS Collaboration},%
pdftitle={Search for a Higgs boson decaying into  a Z and a photon in pp collisions at sqrt(s) = 7 and 8 TeV},%
pdfsubject={CMS},%
pdfkeywords={CMS, physics, Higgs}}

\maketitle 
\section{Introduction}
\label{sec:intro}

\begin{figure}[hbtpH]
   \begin{center}
     \includegraphics[width=\cmsFigWidthX]{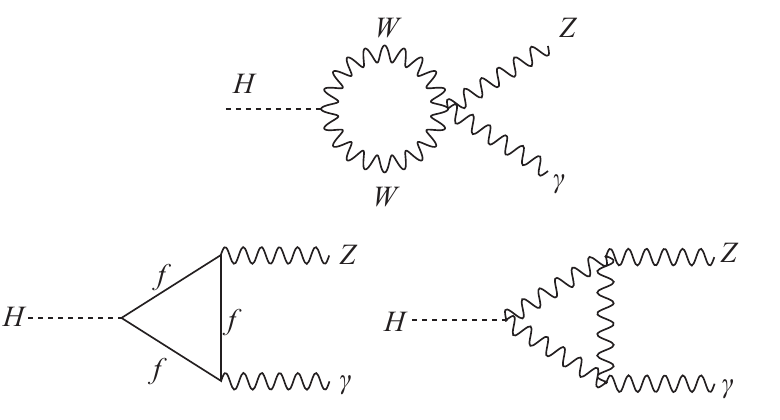}
     \caption{Diagrams contributing to $\Gamma_{\Z\gamma}$.
     \label{fig:fey}
 }
   \end{center}
\end{figure}

The observation of a new resonance decaying to two bosons and  with  decay modes and properties
consistent with those of the standard model\,(SM) Higgs boson has been reported by the ATLAS~\cite{atlas_h,Aad:2013xqa}
 and CMS~\cite{cms_h,long} collaborations. Measurements of the basic properties
of this resonance, such as the mass~\cite{eleReg1b} and the
coupling strength to vector bosons and fermions~\cite{long,cms_h,atlas_h,Aad:2013wqa,Aad:2013xqa}, have been reported.
Within the SM, the partial width for the $\PH\to\Z\gamma$ decay channel ($\Gamma_{\Z\gamma}$)
is rather small, resulting in a branching fraction between  0.11\% and 0.25\% in  the
120--160\GeV~\cite{Cahn:1978nz,Bergstrom:1985hp} mass range.
A measurement of $\Gamma_{\Z\gamma}$ provides important information
on the underlying dynamics of the Higgs sector
because it is induced by loops of heavy charged particles, just as for the
$\PH\to\gamma\gamma$ decay channel.
The contributing diagrams to $\Gamma_{\Z\gamma}$ are shown in Fig.~\ref{fig:fey}.
$\Gamma_{\Z\gamma}$ is sensitive to physics beyond the SM,
and could be  substantially modified by new charged particles without affecting the gluon-gluon fusion Higgs boson
production cross section~\cite{Carena:2012xa},
such as derived from an extended Higgs sector~\cite{Chiang:2012qz}, or by the presence of new scalars~\cite{Gainer:2011aa,Chen:2013vi}.

This paper describes the first search for a   Higgs boson in the $\PH\rightarrow\Z\gamma$ final state
at the LHC in the 120--160\GeV  mass range,
 with the $\Z$ boson decaying into an electron or a muon pair.
This is a
clean final-state topology with an effective mass peak  resolution of about 1-3\%.
To improve the sensitivity of the search,
the selected dilepton-plus-photon events are subdivided into classes according to their mass
resolution and the signal-to-background ratio, for both the electron and muon channels.
The dominant backgrounds consist of the irreducible contribution from
the SM $\Z\gamma$  production, and the reducible backgrounds from final-state-radiation in Drell--Yan  or $\Z$ decays,
and $\Z$ plus jets, where a jet  is misidentified as a photon.
A previous search for $\PH\to\Z\gamma$  has been performed  at the Tevatron for masses above 140\GeV~\cite{Abazov:2008wg}.

Results are based on data samples recorded by the CMS experiment
corresponding to  integrated luminosities of 5.0\fbinv  at 7\TeV and 19.6\fbinv  at 8\TeV in proton-proton collisions.

\section{The CMS detector}
\label{sec:cms}
A detailed description of the CMS detector can be found in Ref.~\cite{ref:cms}.
The central feature of the CMS apparatus is a superconducting solenoid, 13\unit{m} in
length and 6\unit{m} in diameter, which provides an axial magnetic field of 3.8\unit{T}.
Within the field volume there are several particle detection systems. Charged-particle trajectories
are measured by silicon pixel and strip trackers, covering $0\le \phi \le 2\pi$ in azimuth
and $\abs{\eta}  < 2.5$  in pseudorapidity,  where $\eta$ is defined as $-\ln[\tan (\theta/2)]$
and $\theta$ is the polar angle of the trajectory of the particle with respect to the
counterclockwise proton beam direction.
A lead tungstate crystal electromagnetic calorimeter is distributed in a barrel region $\abs{\eta} < 1.48$ and two endcaps
that extend up to $\abs{\eta} = 3$. A brass and scintillator hadron calorimeter surround the tracking volume and cover the region
$\abs{\eta} < 3$.  Iron forward calorimeters with quartz fibers, read out by photomultipliers, extend the calorimeter coverage up to  $\abs{\eta} = 5$.
They  provide   measurements of  the energy of photons, electrons, and hadron jets.
A lead and silicon-strip preshower detector is located in front of the endcap electromagnetic
calorimeter. Muons are identified and measured in gas-ionization
detectors embedded in the steel return-yoke outside the solenoid. The detector is nearly hermetic,
allowing energy balance measurements in the plane transverse to the beam direction. A two-tier
trigger system selects proton-proton collision events of interest for use in physics
analysis.

\section{Event selection}
\label{sec:preselection}

Events with two opposite-sign, same-flavor leptons ($\Pe$ or $\mu$)
and a photon are selected: $\Pep\Pem\gamma$, $\mu^+\mu^-\gamma$.
All particles must be isolated and have  transverse momentum, {$\pt$},  greater than 20  (10)\GeV
for the highest-$\pt$ (next to highest-$\pt$) lepton and 15\GeV for the photon. The electrons\,(muons)
and the photon must have $\abs{\eta} < 2.5\:(2.4)$.
Photons in the barrel-endcap transition region $1.44 < \abs{\eta} < 1.57$ of the electromagnetic
calorimeter are excluded.

Events are required to pass at least one of the dielectron or dimuon
triggers.  The trigger efficiency for events containing two leptons satisfying the
offline event selection requirements
are  measured to be between 60\% and 98\% for the $\Pe\Pe\gamma$ channel depending on the electron
transverse momenta and  91\% for the $\mu\mu\gamma$  channel.

Events are required to have at least one  primary vertex,
with the  reconstructed longitudinal  position ($z$)  within 24\unit{cm} of the geometric
center of the detector and the  transverse position ($x$-$y$) within 2\unit{cm} of the beam interaction region. In the case of
multiple reconstructed  vertices associated with  additional interactions (pileup), the one with the highest  scalar sum of the ${\pt^2}$
of its associated tracks is chosen as the primary vertex. The
leptons  are required to originate
at the same primary vertex.
Electron\,(muon) tracks are required to have the transverse and longitudinal
impact parameters with respect to the primary vertex to be smaller than 2\,(2)\unit{mm} and 2\,(5)\unit{mm}, respectively.

The observables used in the photon selection are: isolation
variables based on the particle-flow (PF) algorithm~\cite{ref:pf}, the
ratio of the energy in the hadron calorimeter  towers behind
the supercluster
to the electromagnetic energy in the supercluster, the transverse width of  the
electromagnetic shower, and a pixel  tracker  veto to avoid misidentifying an electron as a photon.
In the barrel region, superclusters are formed from five crystal strips in $\eta$,
centered on the most energetic crystal, and have a variable extension in $\phi$~\cite{long,hgg}.
In the endcaps, where the crystals are arranged according to an $x$-$y$ rather than an $\eta$-$\phi$
geometry, matrices of 5$\times$5 crystals around the most energetic crystals are merged if they lie
within a narrow $\phi$ road.
The efficiency of the photon identification is measured from $\Z \to \Pe\Pe$ data  using a
tag-and-probe technique~\cite{cite:tagandprobe} by treating the electrons as photons~\cite{long}, and found to be 76\% (88\%)  at a transverse energy  of 15 (above 50)\GeV. These efficiencies include the effects of
the pixel tracker  veto, estimated with $\Z\rightarrow\mu\mu\gamma$ data,
where the photon is produced via final-state radiation.

The electron selection criteria are optimized for background rejection
using a multivariate approach, while maintaining a combined
identification and isolation efficiency of approximately 60\% at low
transverse momentum (10\GeV) and 90\% at high transverse
momentum (50\GeV) for electrons from $\PW$ or $\Z$ boson decays as described in~\cite{Baffioni:2006cd}.
The training of the multivariate electron reconstruction is performed using simulated events, while the
performance is validated using data.  In addition, the electron energy resolution is improved by using a multivariate regression
technique~\cite{eleReg1b}  resulting in improvements of 10\% and 30\% in the
mass resolution for $\Z\to \Pe\Pe$  events over the standard CMS electron
reconstruction in the barrel and endcap, respectively, as described in~\cite{Baffioni:2006cd}.

Muon candidates are reconstructed with a global trajectory fit using
hits in the tracker and the muon system. Muon combined identification and isolation
efficiencies of better than 95\% have been maintained~\cite{eleReg1b,long} after improving the pileup corrections with respect
to those used in~\cite{Chatrchyan:2012xi} for low luminosity data.

Electrons and muons from $\Z$ boson decays are expected to be isolated from other particles.
A cone of size $\DR \equiv \sqrt{\smash[b]{(\Delta\eta)^2 + (\Delta \phi)^2}} = 0.4$  is
constructed around the momentum direction of each considered
lepton candidate~\cite{Baffioni:2006cd,Chatrchyan:2012xi}.
The relative isolation of the lepton is quantified by summing the transverse momentum of
all photons, charged and neutral hadrons PF objects within this cone, excluding the lepton
and charged particles associated with the pileup vertices, and then dividing by the lepton transverse momentum.
The resulting quantity, corrected
for additional underlying event activity due to pileup events~\cite{long}, is required to be less than
0.4 for both $\Z\to\Pep\Pem$ and $\Z\to\mu^+\mu^-$.  This requirement
rejects misidentified leptons and  background arising  from hadronic jets.
Similarly, to reduce the background from misidentified jets in the photon reconstruction,
photon clusters are required to be isolated from other particles within a cone size of $\DR = 0.3$.
Their absolute isolation from charged hadrons, neutral hadrons and photons is
required to be smaller than 1.5 (1.2), 1.0 (1.5), and 0.7 (1.0)\GeV,
respectively,  for photons in
the barrel (endcap) region. These requirements are  applied  after correcting for pileup effects.

The $\ell^+\ell^-$  pair invariant mass is required to be greater than 50\GeV.
No upper dilepton mass condition is needed as the events are found to be dominated by processes containing a $\Z$ boson.
The minimum dilepton mass requirement rejects contributions from $\Pp\Pp\to\gamma\gamma^*$ and  $\PH\to\gamma\gamma^*$ where an internal conversion
of the photon produces a dilepton pair~\cite{cite:higgsDalitz}.
If two dilepton pairs are present, the one with the invariant mass closest to the $\Z$ boson mass is taken.
The final set of requirements combines the information from the photon and the leptons:
(i)\,the invariant mass of the $\ell^+\ell^-\gamma$ system, $m_{\ell\ell\gamma}$,
 is required to be between 100 and 190\GeV;
(ii)\,the ratio of the photon transverse energy to $m_{\ell\ell\gamma}$ must be greater than 15/110;
this requirement suppresses backgrounds due to misidentification of  photons
 without significant loss in signal sensitivity and without introducing a bias
in the $m_{\ell\ell\gamma}$ spectrum;
(iii)\,the $\DR$ separation  between each lepton and the photon must be greater than 0.4\,in order to reject events with
final-state radiation (Drell--Yan or $\Z$ decays); and (iv)\,the remaining final-state radiation events
are rejected by requiring the sum of $m_{\ell\ell\gamma}$ and $m_{\ell\ell}$  to be at least 185\GeV.

Jets used in the dijet-tagged event selection defined below
 are built by clustering the PF candidates with the anti-$\kt$
clustering algorithm~\cite{ref:kt}  with distance parameter of 0.5.
Jets
with a significant fraction of energy coming from pileup interactions
or  not
associated with  the primary vertex are rejected~\cite{long}.
The pileup energy in jets is subtracted using the \textsc{FastJet} technique~\cite{pile1,pile2,pile3}.
Calibrated and corrected
jets~\cite{ref:jetscale} are required to have $\ET  > 30$\GeV,
 $\abs{\eta} < 4.7$, and to be separated by  at least 0.5  in $\DR$  from leptons passing the selection
requirements described above.

\begin{figure}[hbt]
  \begin{center}
    \includegraphics[width=\cmsFigWidthX]{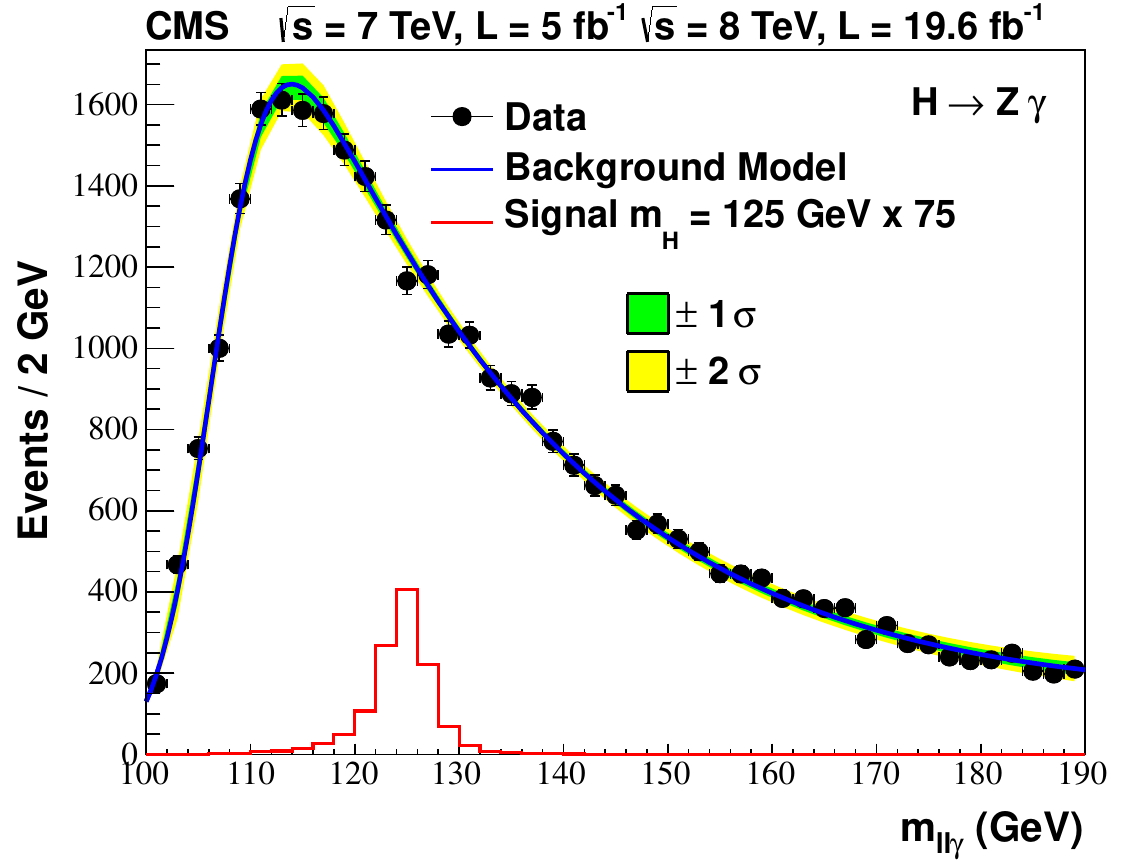}
    \caption{The $m_{\ell\ell\gamma}$ spectrum in the electron and the muon
channels for the 7 and 8\TeV data combined, without weighting by the expected
signal to background ratio of the individual data samples.  Also shown is the expected
signal  due to a 125\GeV standard model Higgs boson, scaled by 75,  and the sum of the individual
fits made to the data for each channel and event class described in section~\ref{sec:classes}.
    The uncertainty band reflects the statistical uncertainty from the fits to the data.
    \label{fig:1}
}
  \end{center}
\end{figure}

The observed yields for the basic event selection described above are listed in Table~\ref{tab:1}.
The total yield for all channels combined is shown in Fig.~\ref{fig:1}.

\begin{table*}[htb]
\begin{center}
\topcaption{
Observed and expected event yields for a 125\GeV
SM Higgs boson.   \label{tab:1}}
\begin{tabular}{l|D{.}{.}{2.1}D{.}{.}{5.0}D{.}{.}{1.1}}
\hline
Sample  & \multicolumn{1}{c}{Integrated} & \multicolumn{1}{c}{Observed event} &  \multicolumn{1}{c}{Expected number  of} \\
        & \multicolumn{1}{c}{luminosity} & \multicolumn{1}{c}{yield for}      &  \multicolumn{1}{c}{signal events for} \\
        & \multicolumn{1}{c}{(fb$^{-1}$)}& \multicolumn{1}{c}{100 $< m_{\ell\ell\gamma} <$190\GeV}       & \multicolumn{1}{c}{$m_\PH=125$\GeV} \\
\hline
2011 $\Pe\Pe$& 5.0  & 2353  & 1.2 \\
2011 $\mu\mu$& 5.1  & 2848  & 1.4  \\
2012 $\Pe\Pe$& 19.6  & 12899 & 6.3 \\
2012 $\mu\mu$& 19.6  & 13860  & 7.0 \\
\end{tabular}
\end{center}
\end{table*}

\section{Event classes}
\label{sec:classes}

The sensitivity of the search is enhanced by 20--40\% by dividing the selected events into  mutually-exclusive classes
according to the expected mass resolution and the signal-to-background ratio,
and then  combining the results from  each class.

\begin{table*}[htbH]
\begin{center}
\topcaption{Definition of the four untagged event classes and the dijet-tagged event class, the fraction of selected events
for a signal with  $m_\PH = 125\GeV$  produced by gluon-gluon  fusion at $\sqrt{s}=8$\TeV, and data in a narrow
region in the $m_{\ell\ell\gamma}$ phase space  centered at 125\GeV. The width of this region is equal to two times the effective standard deviation ($\sigma_\text{eff}$), where $\sigma_\text{eff}$ is defined as half-the-width of the narrowest window containing 68.3\% of the distribution.
The expected full width at  half maximum (FWHM) for the signal is also listed.
\label{tab:2}}
{\scriptsize
\begin{tabular}{ccc}

           & {$\Pep\Pem\gamma$} &  {$\mu^+\mu^-\gamma$} \\ \cline{2-3}
           & \multicolumn{2}{c}{Event class 1}  \\\cline{2-3}
           & {Photon $0<\abs{\eta}<1.44$} &  {Photon $0<\abs{\eta}<1.44$}\\
           & {Both electrons $0<\abs{\eta}<1.44$} & {Both muons $0<\abs{\eta}<2.1$}\\
           & {} & {and one muon $0<\abs{\eta}<0.9$}\\
           & {$\RNINE > 0.94$} & {$\RNINE > 0.94$}\\
\cline{2-3}
Data & {17\%} &{20\%}\\
Signal & {29\%} & {33\%}\\
$\sigma_\text{eff}$ (\GeVns{})& {1.9\GeV}& {1.6\GeV}\\
FWHM (\GeVns{})           & {4.5\GeV}& {3.7\GeV}\\
\cline{2-3}
           & \multicolumn{2}{c}{Event class 2}  \\\cline{2-3}
           & {Photon $0<\abs{\eta}<1.44$} &  {Photon $0<\abs{\eta}<1.44$}\\
           & {Both electrons $0<\abs{\eta}<1.44$} & {Both muons $0<\abs{\eta}<2.1$}\\
           & {} & {and one muon $0<\abs{\eta}<0.9$}\\
           & {$\RNINE < 0.94$} & {$\RNINE < 0.94$}\\
\cline{2-3}
Data & {26\%} &{31\%}\\
Signal & {27\%} & {30\%}\\
$\sigma_\text{eff}$ (\GeVns{})& {2.1\GeV}& {1.9\GeV}\\
FWHM (\GeVns{})          & {5.0\GeV}& {4.6\GeV}\\
\cline{2-3}
 & \multicolumn{2}{c}{Event class 3}  \\\cline{2-3}
           & {Photon $0<\abs{\eta}<1.44$} &  {Photon $0<\abs{\eta}<1.44$}\\
           & {At least one electron $1.44<\abs{\eta}<2.5$} & {Both muons in $\abs{\eta}>0.9$}\\
           & {                                             } & {or one muon in $2.1<\abs{\eta}<2.4$}\\
           & {No requirement on $\RNINE$} & {No requirement on $\RNINE$}\\
\cline{2-3}
Data & {26\%} &{20\%}\\
Signal & {23\%} & {18\%}\\
$\sigma_\text{eff}$ (\GeVns{})& {3.1\GeV}& {2.1\GeV}\\
FWHM (\GeVns{})          & {7.3\GeV}& {5.0\GeV}\\
\cline{2-3}
           & \multicolumn{2}{c}{Event class 4}  \\\cline{2-3}
           & {Photon $1.57<\abs{\eta}<2.5$} &  {Photon $1.57<\abs{\eta}<2.5$}\\
           & {Both electrons $0<\abs{\eta}<2.5$} & {Both muons $0<\abs{\eta}<2.4$}\\
           & {No requirement on $\RNINE$} & {No requirement on $\RNINE$}\\
\cline{2-3}
Data & {31\%} &  {29\%}\\
Signal &   {19\%} & {17\%}\\
$\sigma_\text{eff}$ (\GeVns{})&   {3.3\GeV}&   {3.2\GeV}\\
FWHM (\GeVns{})         &   {7.8\GeV}&   {7.5\GeV}\\
\cline{2-3}
           & \multicolumn{2}{c}{Dijet tagged class}  \\\cline{2-3}
           & {Photon $0<\abs{\eta}<2.5$} &  {Photon $0<\abs{\eta}<2.5$}\\
           & {Both electrons $0<\abs{\eta}<2.5$} & {Both muons $0<\abs{\eta}<2.4$}\\
           & {No requirement on $\RNINE$} & {No requirement on $\RNINE$}\\
\cline{2-3}
Data & {0.1\%} &  {0.2\%}\\
Signal &   {1.8\%} & {1.7\%}\\
$\sigma_\text{eff}$ (\GeVns{})&   {2.6\GeV}&   {2.2\GeV}\\
FWHM (\GeVns{})          &   {4.4\GeV}&   {3.8\GeV}\\
\cline{2-3}
\end{tabular}
}
\end{center}
\end{table*}

As shown in Table~\ref{tab:2},
a significant fraction of the signal events are expected to have both leptons and the photon
in the barrel, while less than a  fifth of the signal events are expected to have a photon in the endcap.
This is in contrast with the background, where around one  third of the events are expected to have a photon in the endcap.
In addition, events where the photon does not convert into an ${\Pep\Pem}$ pair have less background and better resolution
in $m_\PH$.
For these reasons, the events are classified according to the pseudorapidity of the leptons,
the pseudorapidity  of the photon  and the shower shape of the photon for events with the
two  leptons in the  barrel.
The shower shape of the photon ($\RNINE$) is characterized  by
the energy sum of $3 \times 3$ crystals centered on the
most energetic crystal in the supercluster  divided by the energy of the supercluster.
A requirement of a high value of  $\RNINE> 0.94$ is used to identify unconverted photons.
Using this information, the first four event classes are defined as
shown in Table~\ref{tab:2}.
In these four event classes, the  best signal-to-background ratio is obtained for the event class 1, which
is  composed of
events with both leptons and the  photon in the barrel and high $\RNINE$.

It is possible to define an additional class of events with an expected signal-to-background ratio that is
more than an order of magnitude larger than events in the four classes defined above. This is achieved
by requiring  two forward jets with large pseudorapidity separation, to enhance the selection of
Higgs bosons produced via vector boson fusion.
The dijet-tagged event class  requirements are: (i)\,the difference in pseudorapidity between the two jets is
greater than 3.5; (ii)\,the Zeppenfeld variable~\cite{Rainwater:1996ud}
$\eta_{\Z\gamma} - (\eta_{j_1}+\eta_{j_2})/2$ is less than 2.5; (iii)\,the dijet mass is   greater than
500\GeV; and (iv)\,the  difference in azimuthal angles between the dijet system and the $Z\gamma$
system is greater than 2.4.  The dijet selected events form an exclusive event class.
A 10--15\% increase in sensitivity is obtained by adding this event class.
There is a 20\% contribution  from the gluon-gluon fusion production process in the dijet-tagged  event class.
As shown in  Table~\ref{tab:2}, around 2\% of the expected signal events for a  125\GeV  Higgs boson
 belong to this event class, while less than 0.2\% of the background satisfies the dijet-tagged event class  requirements.

\section{Background and signal modeling}
\label{sec:background_signal}

Based on simulated events, the
dominant background in  untagged events is expected to be  due to  initial-state-radiation
SM $\Z\gamma$ production.  The  background fraction due to
final-state-radiation in Drell--Yan or Z decays is only 5\%, while for some event
classes the contribution from Z plus jets can be as large as 40\%.  This is in contrast
to dijet-tagged events, where it is found that the background due to
Z plus  jets is slightly higher than from $\Z\gamma$.

\begin{figure*}[hbt]
  \begin{center}
    \includegraphics[width=\cmsFigWidth]{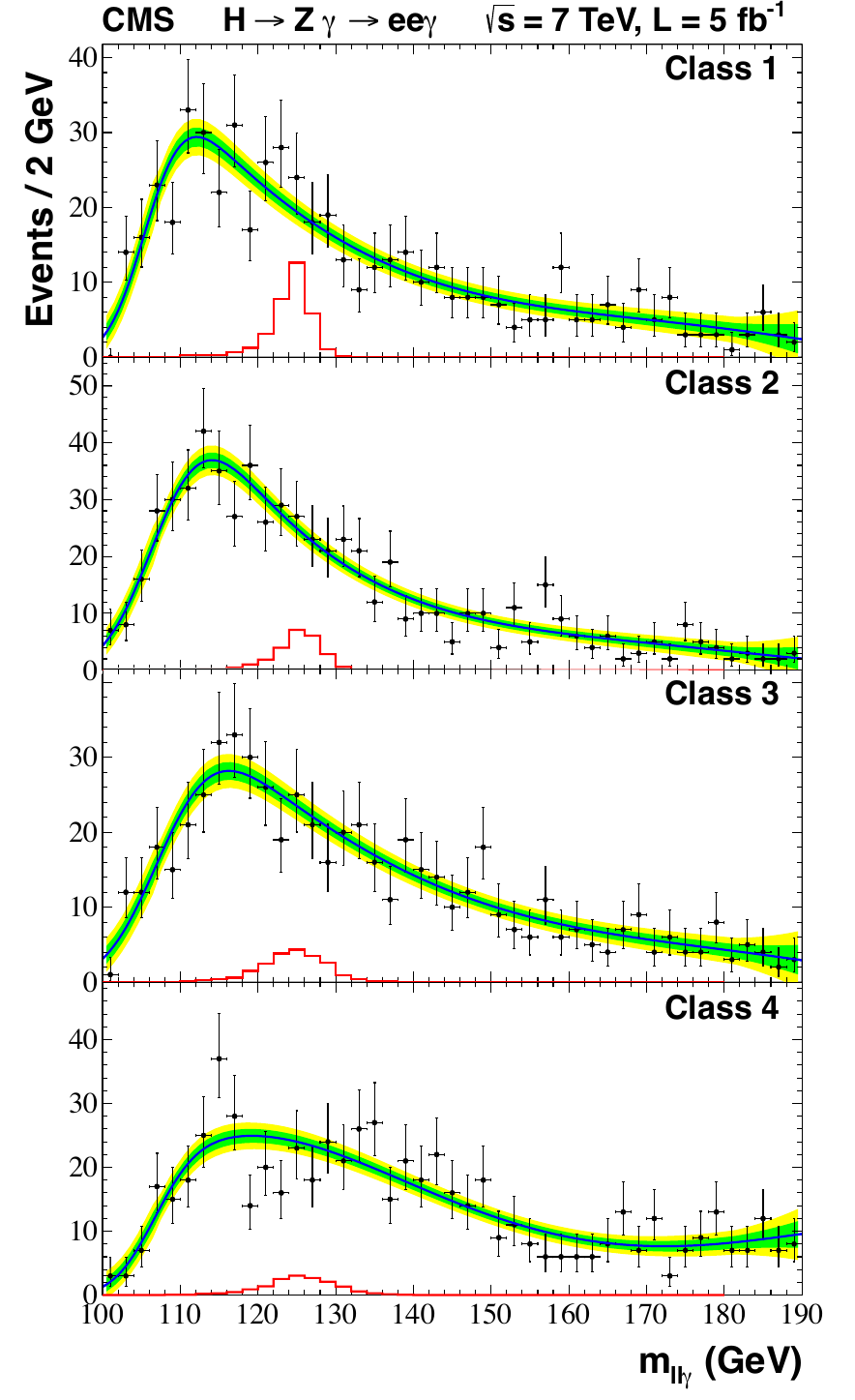}
    \includegraphics[width=\cmsFigWidth]{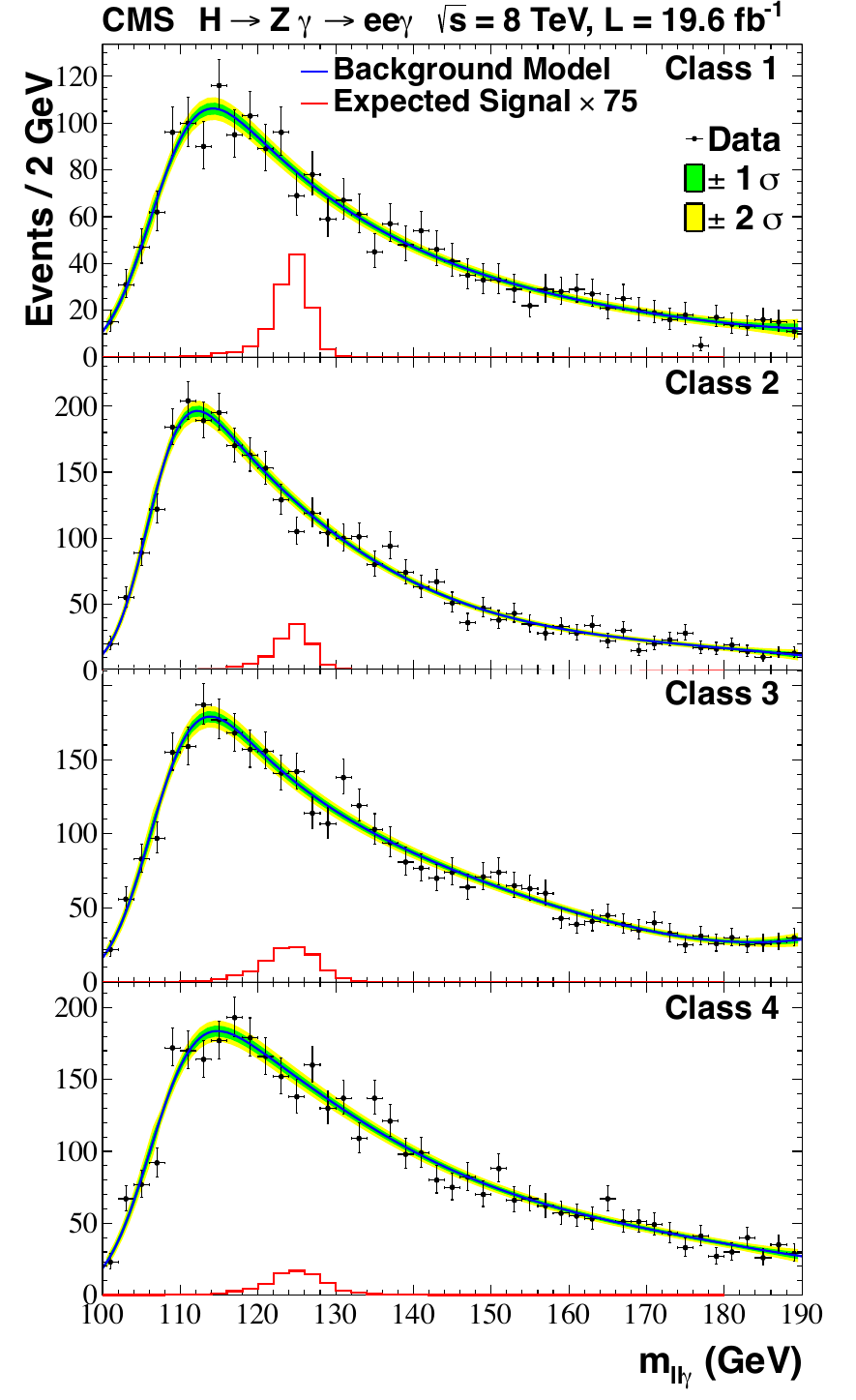}
    \caption{The background model fit to the $m_{{\Pe\Pe}\gamma}$  distributions for event classes  1--4 for the two data samples.
    The statistical uncertainty bands shown are computed from a  fit to data. Also shown is the expected signal due to a 125\GeV standard model Higgs boson, scaled by 75.
    \label{fig:3el}}
  \end{center}
\end{figure*}

\begin{figure*}[hbt]
  \begin{center}
     \includegraphics[width=\cmsFigWidth]{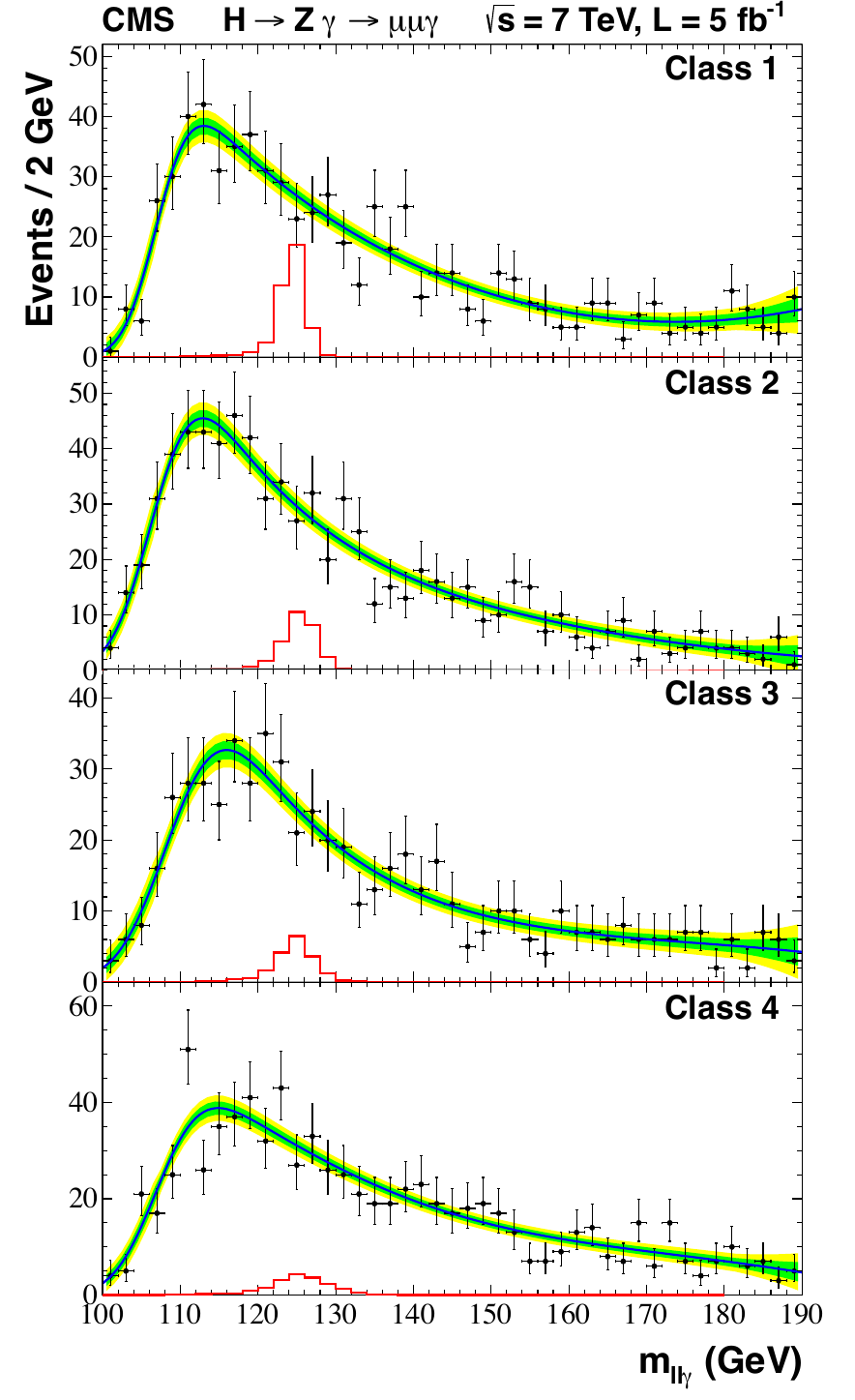}
     \includegraphics[width=\cmsFigWidth]{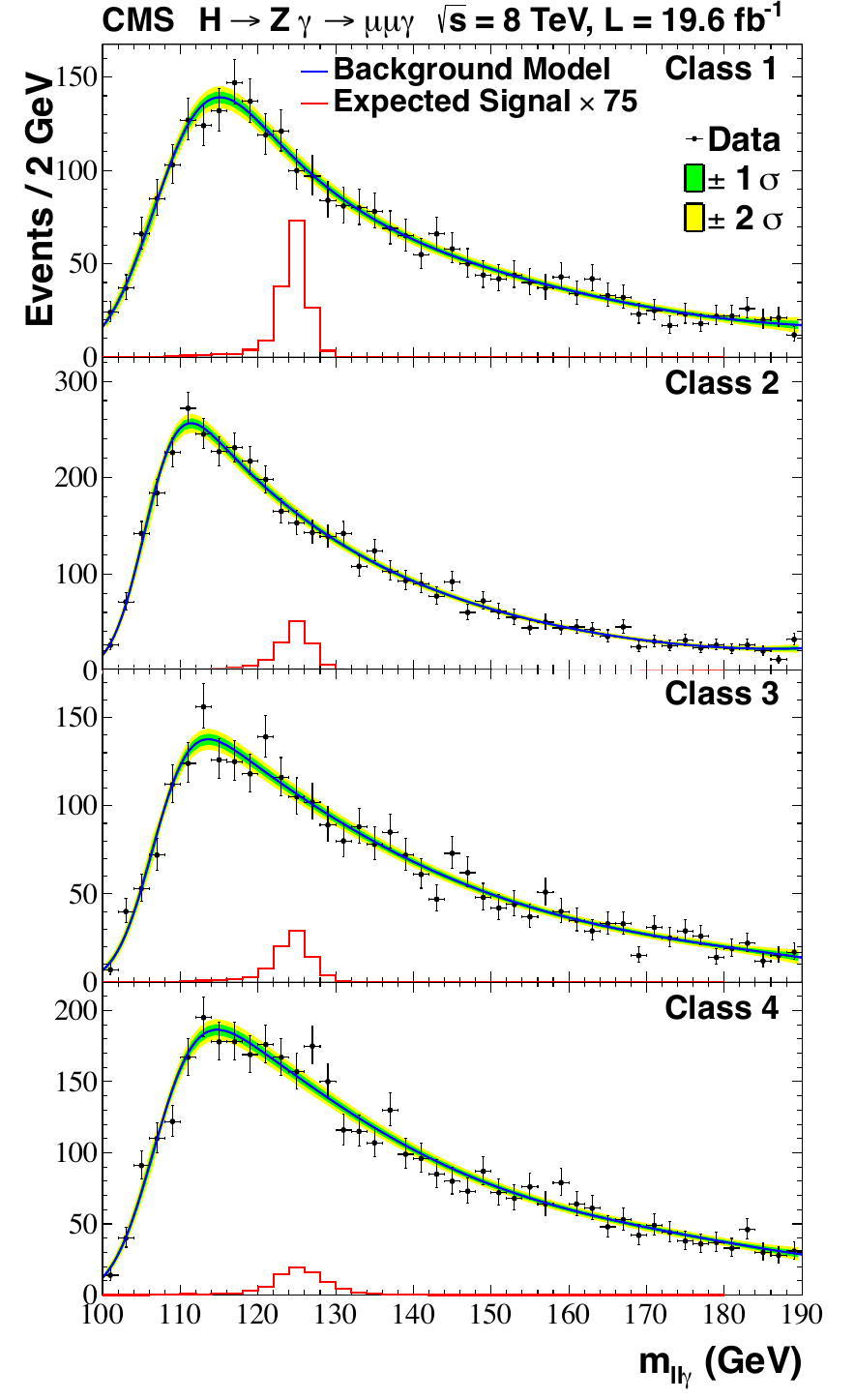}
    \caption{The background model fit to the  $m_{\mu\mu\gamma}$ distributions for event classes  1--4 for the two data samples.
    The statistical uncertainty bands shown are computed from the data fit. Also shown is the expected signal due to a 125\GeV standard model Higgs boson, scaled by 75.
    \label{fig:3mu}}
  \end{center}
\end{figure*}

\begin{figure*}[hbt]
  \begin{center}
    \includegraphics[width=0.325\textwidth]{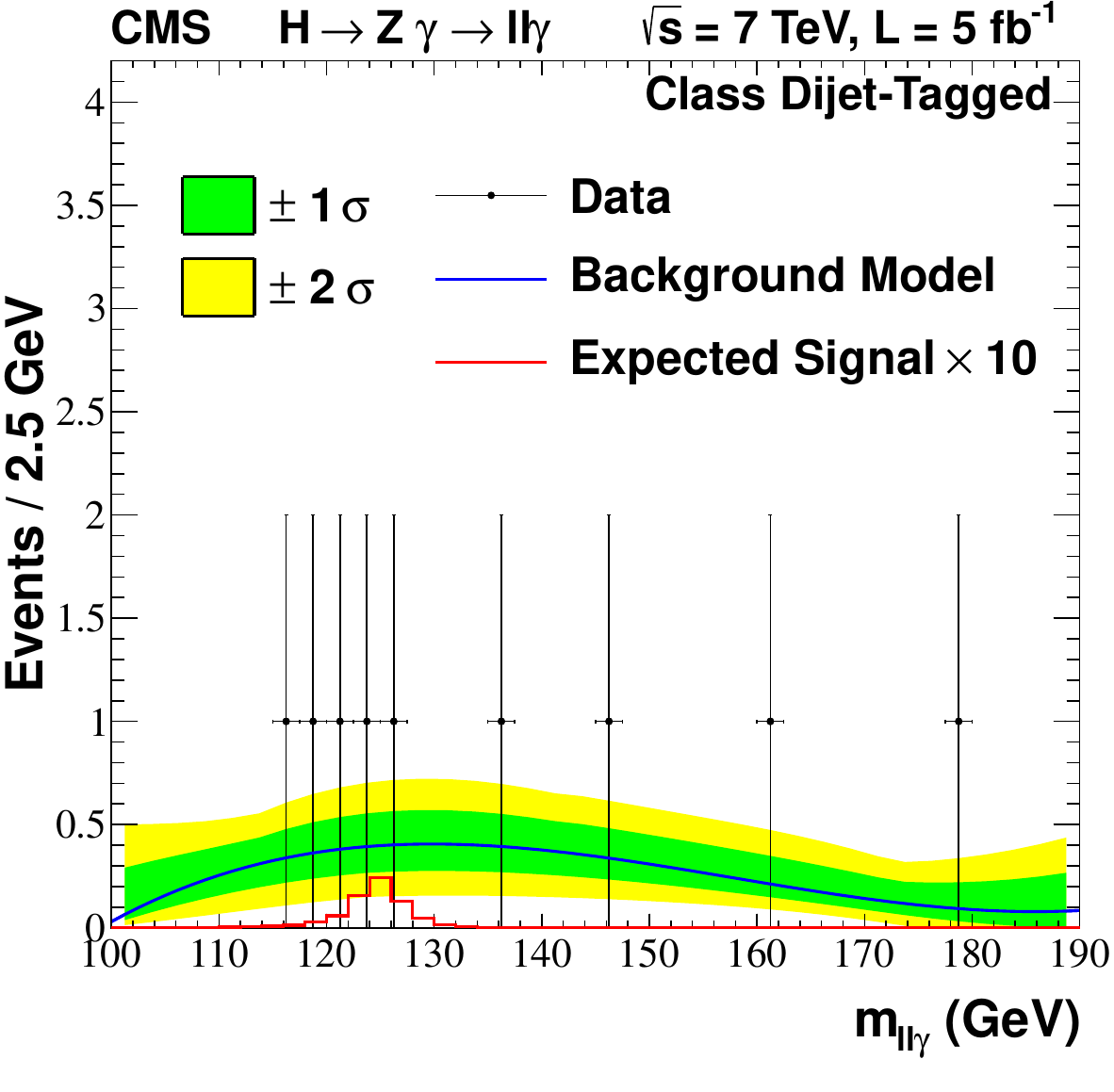}
    \includegraphics[width=0.325\textwidth]{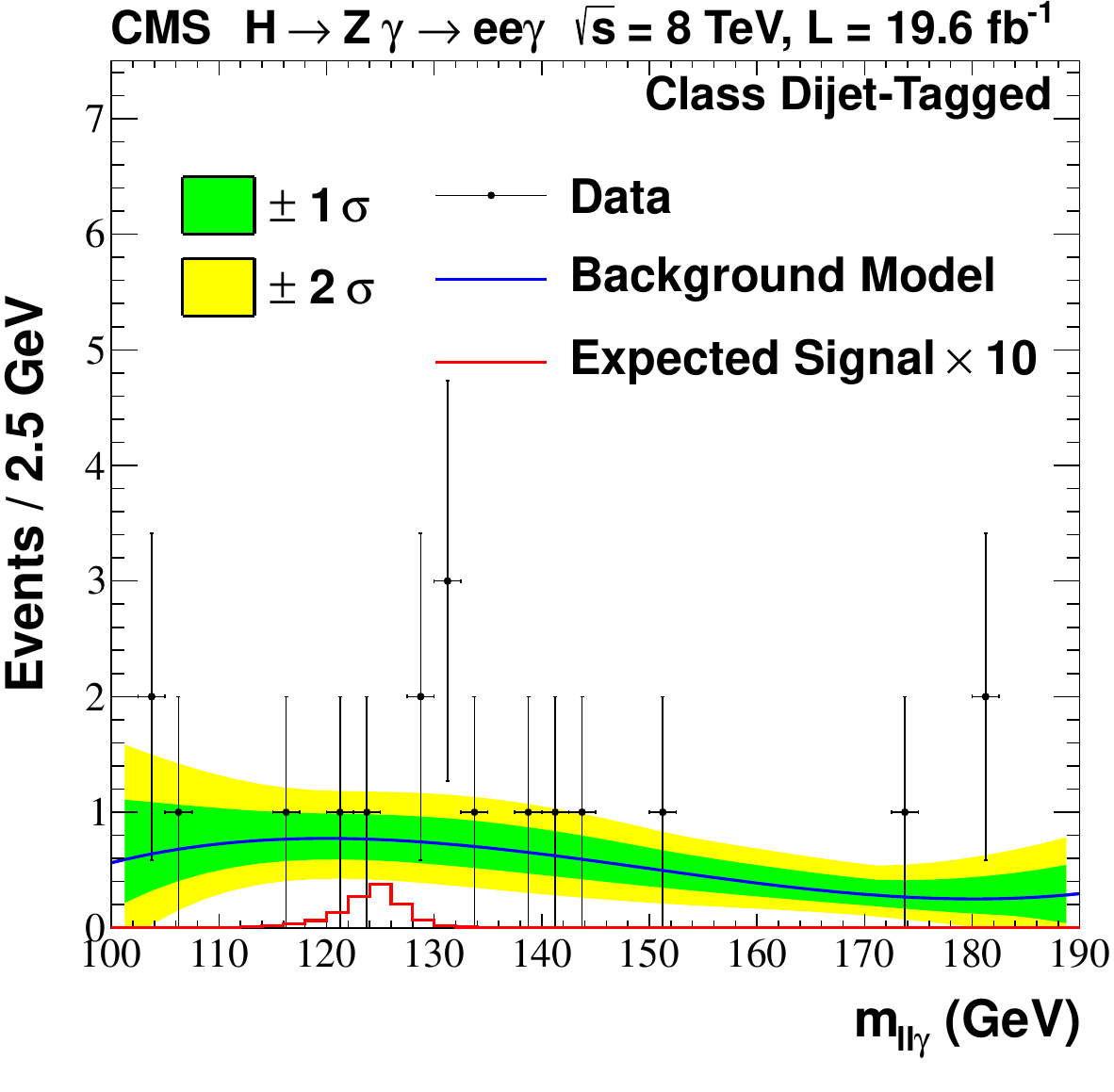}
    \includegraphics[width=0.325\textwidth]{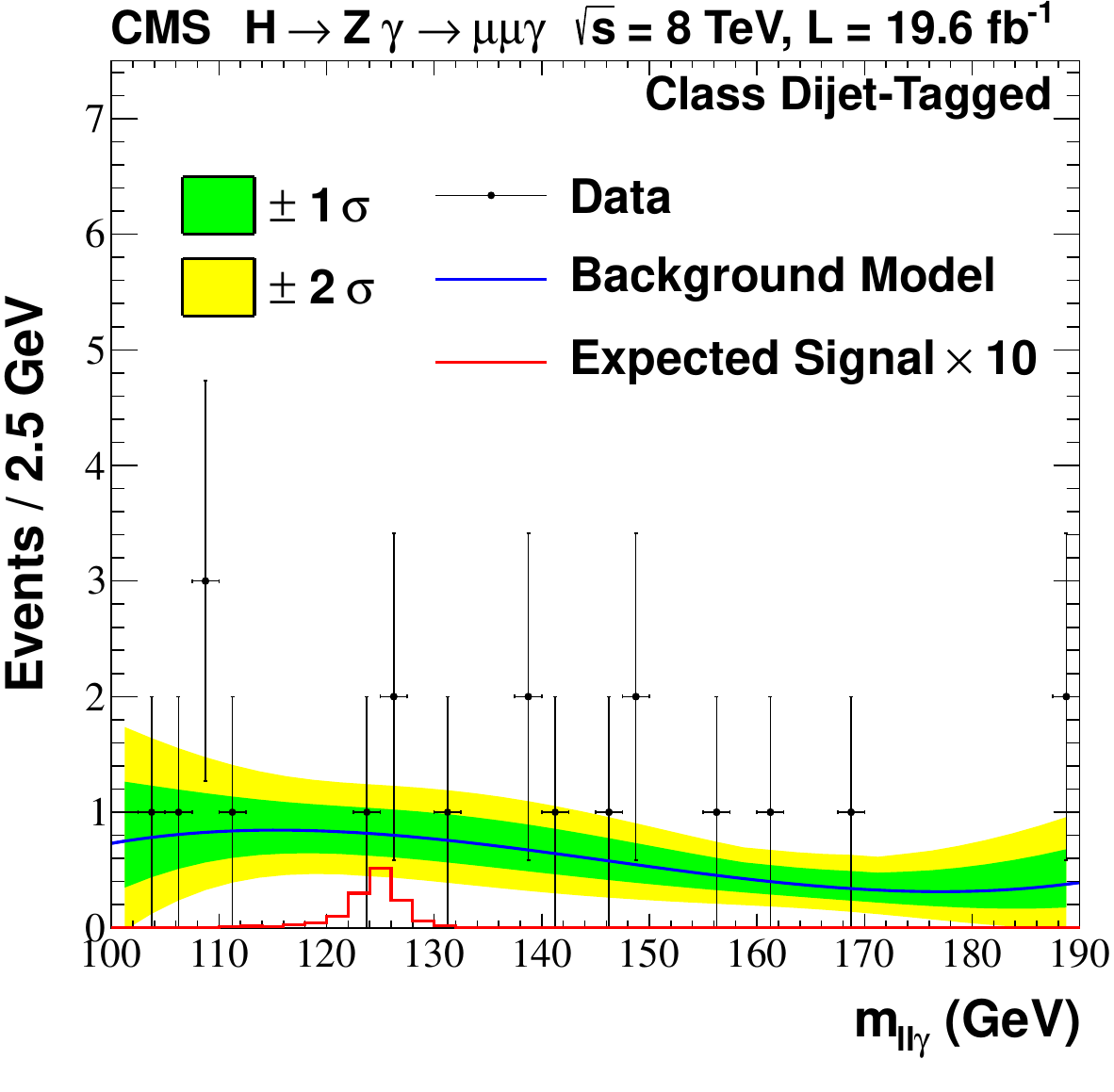}
    \caption{The $m_{\ell\ell\gamma}$ spectrum in the electron and the muon
channels combined (separately) for the 7 (8)\TeV data for the dijet-tagged event class.  The  expected
signal from a 125\GeV standard model Higgs boson has been scaled by a factor of 10.
    \label{fig:vbf}
}
  \end{center}
\end{figure*}

The background model is obtained by fitting the observed $\ell\ell\gamma$ mass distributions
for each of the five event classes
in the electron and the muon channels  at a center-of-mass energy of 7 and 8\TeV separately.
Because of the limited number of events at 7\TeV for
the dijet-tagged event class, the electron and muon channels are combined for this sample.
The fitting is unbinned and is performed over the 100--190\GeV mass range.
The  $m_{\ell\ell\gamma}$  distribution peaks around 110--115\GeV, with a steeply rising turn-on to the left and a
gradually falling tail to the right. These characteristics are fitted to the convolution of a Gaussian with a
step function multiplied by a polynomial. The mean of the Gaussian is fixed to zero in the convolution
and the step position and the width of the Gaussian are left floating in the fit.
The background fits based on the $m_{\ell\ell\gamma}$ data distributions for the  electron
and muon  channels in the untagged event classes are shown in Fig.~\ref{fig:3el} and Fig.~\ref{fig:3mu},
while Fig.~\ref{fig:vbf} shows the dijet-tagged class.
The quality of the fits is good,  with  reduced $\chi^2$
 between 0.49 and 1.8 for the untagged event classes and 0.16 and 0.28 for the dijet-tagged classes.
Even though the  $\PH\rightarrow\Z\gamma$ search is limited to the mass range where the branching fraction is expected to be at least  0.1\% (i.e. 120--160\GeV),  the wide $m_{\ell\ell\gamma}$ fitting range in the background modeling  is found to be needed using the bias studies described below.

The potential bias on the background measurement is studied by using
pseudo-data generated from background-only fits to the observed $m_{\ell\ell\gamma}$
spectrum.  These pseudo-data sets are fitted to a signal combined with   a polynomial background model.
The results of these fits are used to determine an appropriate degree
of polynomial model for background, such that the bias introduced on the
limit of the signal strength measurement is smaller than a fifth of the
background statistical uncertainty.
This is the same method used in the search for  the Higgs boson decaying to $\gamma\gamma$ and
described in detail in~\cite{long}.
A third-order polynomial  is chosen for the dijet-tagged event class, a
 fourth-order polynomial is chosen to fit
the event classes where both leptons and the photon are in the barrel, while a fifth-order
polynomial is chosen to fit the event classes
where  the photon and at least one lepton are in the endcap.


The description of the Higgs boson signal used in the search is obtained from simulated events
produced by the next-to-leading order matrix-element generator
{\POWHEG 1.0}~\cite{cite:powheg1,cite:powheg2} interfaced with {\PYTHIA 6.4}~\cite{cite:pythia}
for the gluon-gluon fusion and vector boson fusion process. The parton distribution functions (PDF)
used to produce these samples is CT10~\cite{Guzzi:2011sv}. Associated production with a vector boson and
associated production with a \ttbar pair are simulated at leading-order using {\PYTHIA 6.4} and the
CTEQ6L~\cite{Pumplin:2002vw} PDF.
The SM Higgs boson cross sections and branching fractions used are taken from Refs.~\cite{lhcwg12,Denner:2011mq}.
The simulated signal events are weighted by taking into account the difference between
data and simulated events so that the distribution of reconstructed vertices,
the trigger efficiencies, the energy and momentum resolution, the energy scale, the reconstruction
efficiency, and the  isolation efficiency
observed in data  are reproduced for all reconstructed objects. An additional small correction is applied to the photons to reproduce the
performance of the $\RNINE$ shower shape variable, by scaling  the shower shape variable to match those observed in the
$\Z\to\mu\mu\gamma$ events~\cite{hgg}.

\section{Results}
\label{sec:results}

A statistical analysis to test the significance of any potential  signal-like excess is performed
 in terms of the local p-value, the probability of observing an excess under the background-only hypothesis.
The local p-value is expressed as a number of standard deviations using the one-sided Gaussian tail
convention.  No significant excess above background is observed, with a maximum excess
of less than two standard deviations in the full mass range.
The data are used to derive upper limits on the Higgs boson production cross
section times the  $\PH\to\Z\gamma$  branching fraction, $\sigma(\Pp\Pp\to\PH)\times\mathcal{B}(\PH\to\Z\gamma)$.
The limits are evaluated using a modified frequentist approach taking the profile
likelihood as a test statistic~\cite{cite:l1,cite:l2,cite:l3}. An unbinned
evaluation of the likelihood over the full mass range of data is used.  In addition, the limit on the inclusive
cross section times the branching fraction is also provided. No theoretical uncertainties on
the production cross sections are included in the latter result.
The uncertainty on the limit is dominated by the size of the data sample and systematic uncertainties have
a negligible impact.

The systematic uncertainty in the limits is only due to the signal description, as the background is
obtained from data and biases are avoided in the fitting procedure.
The uncertainty arises from the uncertainty in the luminosity
measurement (2.2\%\,\cite{cite:lumi-7},\,4.4\%\,\cite{cite:lumi-8}),
the trigger efficiency (0.5--3.5\%), the effects of the
choice of parton distribution functions on the signal cross section (0.3--12.5\%)~\cite{cite:cs1,cite:cs2,cite:cs3,cite:cs4,Ball:2011mu},
the uncertainty in the Higgs boson branching fraction prediction (6.7--9.4\%)~\cite{lhcwg12,Denner:2011mq},
the event pileup modeling for the signal samples (0.4--0.8\%),
the  corrections applied to the simulation to reproduce the performance of the lepton (0.7--1.4\%),
photon (0.5--1.0\%),
and dijet selections (8.8--28.5\%),
event migration caused  by  the requirements on the photon shower shape in the event classification (5.0\%),
the event migration between dijet-tagged and untagged event classes due to the jet energy scale
(5.1--9.8\%), and the signal modeling (1.0--5.0\%).
The uncertainty in the signal modeling takes into account a  potential 5\% contamination from
final-state radiation in the  $\PH\to\mu\mu$ decay, assuming the SM branching fraction.
Based on the fit bias studies performed in the 120--160\GeV mass range,
the uncertainty on the background estimation due to the  chosen functional form is shown to be negligible.

\begin{figure}[hbt]
  \begin{center}
    \includegraphics[width=\cmsFigWidthX]{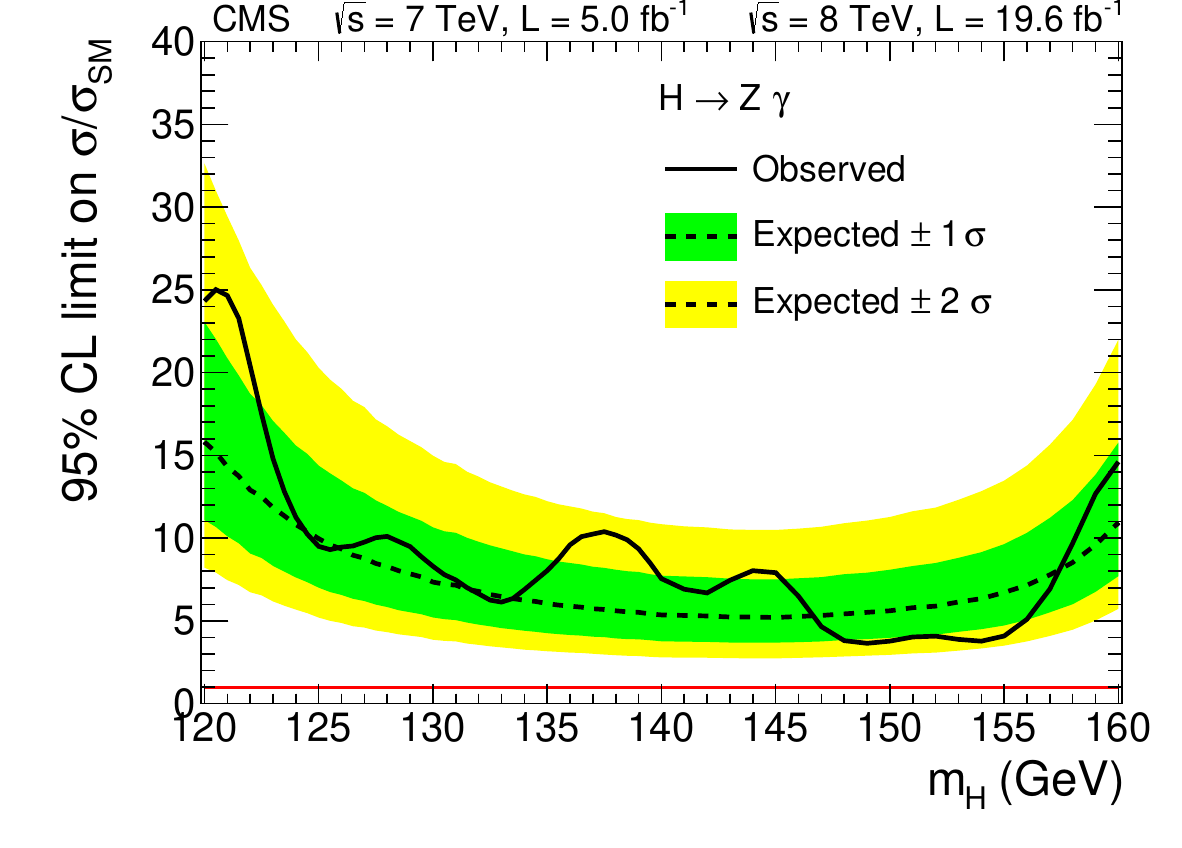}
    \includegraphics[width=\cmsFigWidthX]{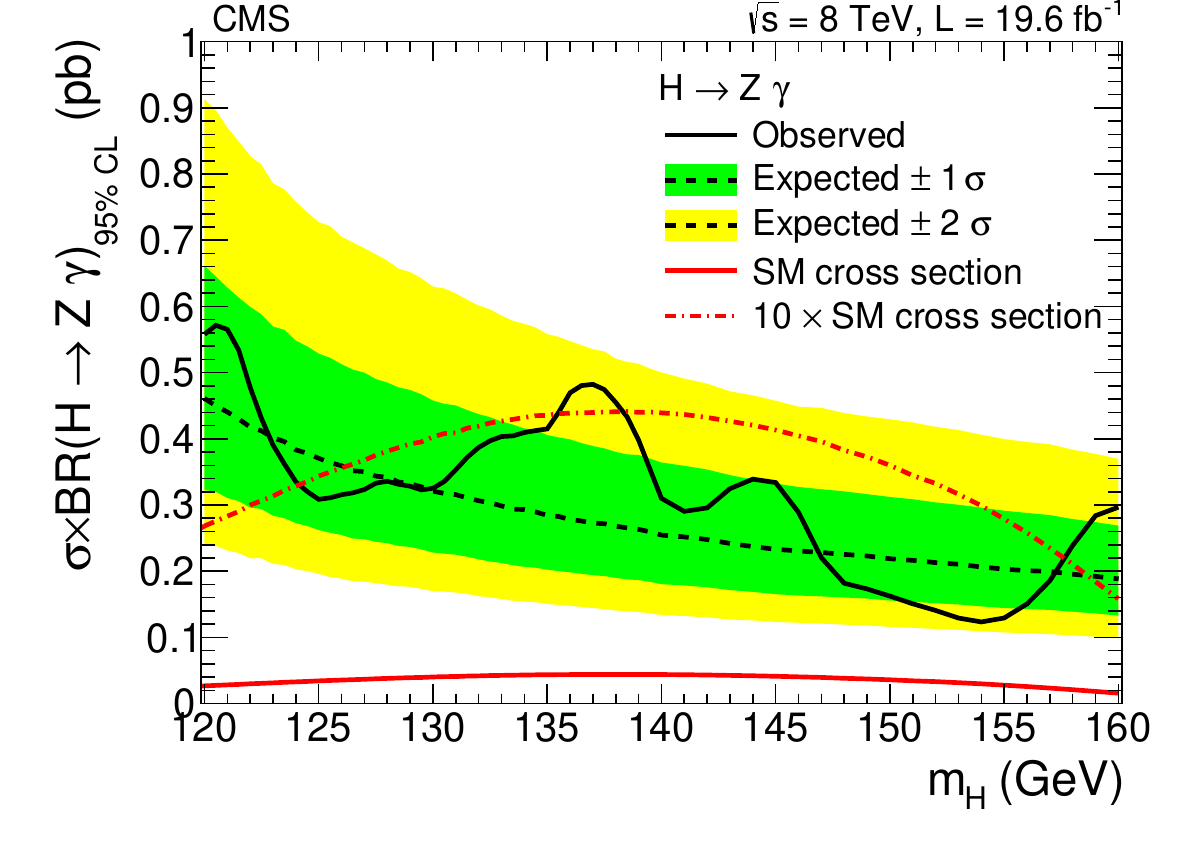}
    \caption{(a) The exclusion limit on the cross section times the branching fraction of a Higgs boson decaying
 into a \Z boson and a photon divided by the SM value. (b) Exclusion limit on the inclusive cross section alone,
where the theoretical uncertainties on the cross section have been excluded in the limit setting.
    \label{fig:4}}
  \end{center}
\end{figure}

The expected and observed limits are shown in Fig.~\ref{fig:4}.
The limits are calculated at 0.5\GeV intervals in the 120--160\GeV mass range.
The expected exclusion limits at 95\% confidence level (CL)
are between 5 and 16 times the SM cross section
 and the observed limit ranges     between about 4 and 25 times the
SM cross section.
The observed  and expected limits for $m_{\ell\ell\gamma}$ at  125\GeV are within one order of magnitude of the SM prediction.
The data excludes models predicting $\sigma(\Pp\Pp\to\PH)\times\mathcal{B}(\PH\to\Z\gamma)$ to be larger than
one order of magnitude of the SM prediction for most of the 125--157\GeV mass range.
Hence, models predicting  significant enhancements for $\Gamma_{\Z\gamma}$~\cite{Coleppa:2012eh}
with respect to the SM expectations due to a pseudoscalar admixture, already strongly disfavoured from the analysis
of the angular distributions of the lepton pairs in $\PH\to\Z\Z$ decays~\cite{eleReg1b},  are now excluded.

\section{Summary}
\label{sec:Summary}

A search has been performed for a Higgs boson decaying into a \Z boson and a photon.
The analysis used  a dataset from proton-proton collisions at a center-of-mass energy of 7 and 8\TeV,
corresponding to an integrated luminosity of 5.0 and 19.6\fbinv, respectively.
No excess above standard model predictions has been found and the first limits on the Higgs boson production
cross section times the  $\PH\to\Z\gamma$  branching fraction at the LHC have been derived.
The expected exclusion limits at 95\% confidence level
are between 5 and 16 times the standard model cross section in the 120--160\GeV mass range
and the observed limit ranges     between about 4 and 25 times the
standard model cross section.
The observed  and expected limits for $m_{\ell\ell\gamma}$ at  125\GeV are within one order of magnitude of the standard model prediction.
Models predicting $\sigma(\Pp\Pp\to\PH)\times\mathcal{B}(\PH\to\Z\gamma)$ to be larger than
one order of magnitude of the standard model prediction, for most of the 125--157\GeV mass range, are excluded.

We congratulate our colleagues in the CERN accelerator departments for the excellent performance of the LHC and thank the technical and administrative staffs at CERN and at other CMS institutes for their contributions to the success of the CMS effort. In addition, we gratefully acknowledge the computing centres and personnel of the Worldwide LHC Computing Grid for delivering so effectively the computing infrastructure essential to our analyses. Finally, we acknowledge the enduring support for the construction and operation of the LHC and the CMS detector provided by the following funding agencies: BMWF and FWF (Austria); FNRS and FWO (Belgium); CNPq, CAPES, FAPERJ, and FAPESP (Brazil); MES (Bulgaria); CERN; CAS, MoST, and NSFC (China); COLCIENCIAS (Colombia); MSES (Croatia); RPF (Cyprus); MoER, SF0690030s09 and ERDF (Estonia); Academy of Finland, MEC, and HIP (Finland); CEA and CNRS/IN2P3 (France); BMBF, DFG, and HGF (Germany); GSRT (Greece); OTKA and NKTH (Hungary); DAE and DST (India); IPM (Iran); SFI (Ireland); INFN (Italy); NRF and WCU (Republic of Korea); LAS (Lithuania); CINVESTAV, CONACYT, SEP, and UASLP-FAI (Mexico); MBIE (New Zealand); PAEC (Pakistan); MSHE and NSC (Poland); FCT (Portugal); JINR (Dubna); MON, RosAtom, RAS and RFBR (Russia); MESTD (Serbia); SEIDI and CPAN (Spain); Swiss Funding Agencies (Switzerland); NSC (Taipei); ThEPCenter, IPST, STAR and NSTDA (Thailand); TUBITAK and TAEK (Turkey); NASU (Ukraine); STFC (United Kingdom); DOE and NSF (USA).

\bibliography{auto_generated}   
\cleardoublepage \appendix\section{The CMS Collaboration \label{app:collab}}\begin{sloppypar}\hyphenpenalty=5000\widowpenalty=500\clubpenalty=5000\textbf{Yerevan Physics Institute,  Yerevan,  Armenia}\\*[0pt]
S.~Chatrchyan, V.~Khachatryan, A.M.~Sirunyan, A.~Tumasyan
\vskip\cmsinstskip
\textbf{Institut f\"{u}r Hochenergiephysik der OeAW,  Wien,  Austria}\\*[0pt]
W.~Adam, T.~Bergauer, M.~Dragicevic, J.~Er\"{o}, C.~Fabjan\cmsAuthorMark{1}, M.~Friedl, R.~Fr\"{u}hwirth\cmsAuthorMark{1}, V.M.~Ghete, N.~H\"{o}rmann, J.~Hrubec, M.~Jeitler\cmsAuthorMark{1}, W.~Kiesenhofer, V.~Kn\"{u}nz, M.~Krammer\cmsAuthorMark{1}, I.~Kr\"{a}tschmer, D.~Liko, I.~Mikulec, D.~Rabady\cmsAuthorMark{2}, B.~Rahbaran, C.~Rohringer, H.~Rohringer, R.~Sch\"{o}fbeck, J.~Strauss, A.~Taurok, W.~Treberer-Treberspurg, W.~Waltenberger, C.-E.~Wulz\cmsAuthorMark{1}
% ...
\vskip\cmsinstskip
% ...
1:~~Also at Vienna University of Technology, Vienna, Austria\\
2:~~Also at CERN, European Organization for Nuclear Research, Geneva, Switzerland\\
% ...

\end{sloppypar}
\end{document}